\documentstyle[aps]{revtex}
\newfont{\bg}{cmr12 scaled\magstep4}
\newcommand{\bigzerol}{\smash{\hbox{\bg 0}}}
\newcommand{\bigzerou}{\smash{\lower1.7ex\hbox{\bg 0}}}
\begin{document}
\title{ 
Spectral Properties of the Two-Dimensional Laplacian  
with a Finite Number of Point Interactions
}
\author{
T. Shigehara$^{1}$, H. Mizoguchi$^{1}$, T. Mishima$^{1}$ 
and Taksu Cheon$^{2}$\\
$^{1}$Department of Information and Computer Sciences, Saitama University \\
Shimo-Okubo, Urawa, Saitama 338 Japan \\
$^{2}$Laboratory of Physics, Kochi University of Technology \\
Tosa Yamada, Kochi 782 Japan 
}
\maketitle

\begin{abstract}
We discuss spectral properties of the Laplacian 
with multiple ($N$) point interactions in two-dimensional bounded regions.  
A mathematically sound formulation for the problem is given 
within the framework of the self-adjoint 
extension of a symmetric (Hermitian) operator in functional 
analysis. 
The eigenvalues of this system are obtained as  
the poles of a transition matrix which has size $N$.  
Closely examining a generic behavior of the eigenvalues 
of the transition matrix as a function of the energy,   
we deduce the general condition under which 
point interactions have a substantial effect on 
statistical properties of the spectrum. 
\end{abstract}

\section{Introduction}

The Laplacian is one of fundamental operators which describe 
real systems and indeed appears in various fields such as 
classical mechanics, electromagnetic dynamics, fluid dynamics, 
quantum mechanics and so on. 
The eigenvalue problem of the Laplacian, 
in particular, has a direct relation to the microscopic world  
where quantum mechanics governs the dynamics. 
The eigenvalues of the Laplacian on a bounded region, 
$\omega_i$, $i=1,\cdots$, exactly correspond to the allowable 
energies of a free particle moving in a microscopic billiard, 
which are discrete in general, and each eigenfunction, 
$\psi_i (\vec{x})$,  
describes the wave function of the corresponding eigenstate, 
the square of which, $| \psi_i(\vec{x}) |^2$, expresses the probability 
such that the particle with energy $\omega_i$ 
exists at a position $\vec{x}$ in the billiard. 

The quantum billiard problem is a natural idealization of 
the particle motion in microscopic bounded regions. 
The one-electron problem in ``quantum dots'' is a possible 
setting which may be used as a single-electron memory, 
a promising computational device in the future. 
It is now possible to actually construct such structures with 
extremely pure semiconductors thanks to the rapid progress 
in the mesoscopic technology. Real systems are, however, not 
free from impurities which affect the particle motion inside. 
In the presence of a small amount of contamination, even a 
single-particle problem becomes unmanageable. 
The modeling of the impurities with point interactions is 
expected to make the problem easy to handle without changing 
essential dynamics. 

Apart from the applicability to physical systems with 
a microscopic or mesoscopic scale, 
the Laplacian with point interactions is an 
attractive object from a viewpoint of dynamical systems. 
The two-dimensional quantum billiard is an appropriate tool 
for examining generic features of dynamical systems because of 
the wide rage of dynamical behaviors, going from the most 
regular (integrable) to the most irregular (chaotic) 
depending on the geometry of its boundary. 
Although no mathematical proof exists, it is widely 
believed that fingerprints of the regular or irregular 
nature of the classical motion can be found in statistical 
properties of the spectrum (eigenvalues and eigenfunctions) 
in the corresponding quantum system \cite{BO89}. 
One of such statistical 
measures is the nearest-neighbor level spacing distribution 
$P(S)$, which is defined such that $P(S)dS$ is the probability 
to find the spacing between any two neighboring energy levels 
in the interval $(S,S+dS)$. 
Integrable systems such as 
circular, elliptic and rectangular billiards obey Poisson 
statistics; 
\begin{eqnarray}
\label{eq1-1}
P(S)=\exp (-S). 
\end{eqnarray}
(It is conventional to take a unit of the energy such that 
the average level spacing is one. Namely, $P(S)$ satisfies 
$\int_0^{\infty}SP(S)dS=1$ as well as $\int_0^{\infty}P(S)dS=1$.) 
On the other hand, 
chaotic systems such as Sinai's billiard and Bunimovich's stadium  
are described by the prediction of the Gaussian orthogonal ensembles;  
\begin{eqnarray}
\label{eq1-2}
P(S)=\frac{\pi S}{2} \exp \left( -\frac{\pi S^2}{4} \right). 
\end{eqnarray}
There are other statistics suitable for a measure of the degree 
of regularity (or irregularity), such as $\Delta_3$-statistic introduced 
by Dyson and Mehta \cite{DM63,ME67}. 
From a viewpoint of dynamical systems, 
the integrable billiards with point interactions inside are classified 
into a category called ``pseudointegrable''. 
The nature of classical motion in the pseudointegrable systems is 
integrable in the sense that a set of 
unstable trajectories are of measure zero in the phase space. 
However, several numerical experiments show that under a certain 
condition, quantization induces the chaotic energy spectra, which might  
be regarded as a counterexample for the correspondence between 
the level statistics and the underlying classical motion. 

One of the main purposes of this paper is to clarify the condition 
under which the statistical properties of quantum spectrum 
are substantially affected by point interactions. 
We spend a large portion of the paper in 
giving a mathematical background to answer this question. 
In spite of its seeming simplicity, 
a careful treatment is demanded for point interactions in quantum 
mechanics in case of the spatial dimension $d \geq 2$. 
Relying on the self-adjoint extension theory in functional analysis, 
we give a general formula suitable for the resolvent (Green's function) 
which describes a particle propagation in 
quantum billiard with point interactions. 
It is hard to proceed further without losing any generality; 
Numerics is, in general, demanded to calculate each eigenvalue 
(pole of the Green's function) and the corresponding eigenfunction. 
By introducing suitable approximations for examining 
statistical properties on spectrum, however, 
we deduce the condition for the strong coupling where 
the effect of point interactions on the quantum spectrum is maximal. 

The paper is organized as follows. 
In Sec.II, the Green's function for the Laplacian with a finite number 
of point interactions is deduced in a mathematically sound manner. 
We discuss the condition for the strong coupling for the case of 
a single interaction in Sec.III, which is extended to the case of 
multiple interactions in Sec.IV. 
The present work is summarized in Sec.V. 

\section{Formulation of Quantum Billiards with Point Interactions}
We start from an empty billiard.
Let us consider a quantum particle of mass $M$ moving 
freely in a bounded region $S$ in two spatial dimension. 
The wave functions are assumed to vanish on its boundary $\partial S$.  
We denote the eigenvalues and corresponding eigenfunctions 
of the system as $\varepsilon _n$ and $\varphi _n$, namely
\begin{eqnarray}
\label{eq2-1}
H_0 \equiv -{\Delta \over {2M}}, \ \ \ 
H_0 \varphi _n({\vec x})
=\varepsilon _n \varphi _n({\vec x}), \ \ \ n=1,\cdots 
\end{eqnarray}
with
\begin{eqnarray}
\label{eq2-2}
 \varphi _n({\vec x})=0
\ \ \   {\rm where} \ \ \  
{\vec x}\in \partial S.
\end{eqnarray}
The Hamiltonian $H_0$ is the kinetic operator in $L^2 (S)$ with 
domain $D(H_0)=H^{2}(S)\cap H^{1}_{0}(S)$ in terms of 
the Sobolev spaces. 
According to the Weyl formula, 
the average level (eigenvalue) density is given by 
\begin{eqnarray}
\label{eq2-3}
\rho_{av} = \frac{MS}{2\pi}
\end{eqnarray}
where we denote the area of $S$ by the same symbol. 
Note that $\rho_{av}$ is independent of the energy $\omega$ 
in two dimension. 
Assuming $\varphi_n({\vec x})$ to be normalized to unity, 
the Green's function of $H_0$ is given by
\begin{eqnarray}
\label{eq2-4}
G^{(0)}({\vec x},{\vec x'};\omega ) \equiv  
( \omega - H_0 )^{-1}({\vec x},{\vec x'}) = \sum\limits_{n=1}^\infty 
{{{\varphi _n({\vec x})\varphi _n({\vec x'})} 
\over {\omega -\varepsilon _n}}}.
\end{eqnarray}

We now place $N$ point interactions at 
${\vec x_1}, ..., {\vec x_N}$.
Naively, one defines the interactions in terms of the 
Dirac's delta function in two dimension; 
\begin{eqnarray}
\label{eq2-5}
H=H_0+\sum\limits_{i=1}^N v_i\ \delta ({\vec x}-{\vec x_i}).
\end{eqnarray}
However, the Hamiltonian $H$ is not mathematically sound  
for spatial dimension $d \geq 2$. 
This can be seen even in case of a single interaction ($N=1$). 
In this case, the eigenvalue equation of $H$ is reduced to 
\begin{eqnarray}
\label{eq2-6}
\sum_{n=1}^{\infty}
\frac{\varphi_n(\vec{x}_1)^2}{\omega-\varepsilon_n}=v_1^{-1}.  
\end{eqnarray}
However, since the average level density is constant  
with respect to the energy, 
the infinite series on the left hand side does not converge 
in two dimension. 
For higher dimensions, 
the average level density has a positive power dependence of 
the energy, which makes the series divergent. 
To handle the divergence, a scheme for regularization and 
renormalization is called for.  One of the most mathematically 
satisfying schemes is given by the self-adjoint extension 
theory of functional analysis \cite{AG88}. 

We first consider in $L^2(S)$ the nonnegative operator 
\begin{eqnarray}
\label{eq2-7}
H_X = - \left. {\Delta \over {2M}} \right|_{C^{\infty}_0 (S-X)}  
\end{eqnarray}
with its closure ${\bar H_X}$ in $L^2{(S)}$,  
where we set $X=\{ {\vec x_1}, ..., {\vec x_N} \}$. 
Namely, we restrict $D(H_0)$ to the functions 
which vanish at the position of the point interactions. 
By using integration by parts, 
it is easy to prove that 
the operator ${\bar H_X}$ is symmetric (Hermitian). 
But it is not self-adjoint. Indeed, the equation 
\begin{eqnarray}
\label{eq2-9}
{\bar H_X}^* \psi(\omega,{\vec x}) = \omega \psi(\omega,{\vec x}), \ \ \ \
\psi \in D({\bar H_X}^*), 
\ \ \ \ Im \ \omega \neq 0,  
\end{eqnarray}
has the N independent solutions \cite{ZO80}
\begin{eqnarray}
\label{eq2-10}
\psi_i (\omega,{\vec x}) =  
G^{(0)}({\vec x},{\vec x_i};\omega ), 
\ \ {\vec x} \in S-X, \ \ i=1,\cdots,N, 
\end{eqnarray}
indicating 
\begin{eqnarray}
\label{eq2-10b}
D(\bar{H}_X^*)=D(\bar{H}_X) \oplus \overline{Ran(\bar{H}_X-\omega)}^{\perp} 
\oplus \overline{Ran(\bar{H}_X-\bar{\omega})}^{\perp} \neq D(\bar{H}_X), \ \ \ 
Im \ \omega \neq 0.   
\end{eqnarray}
Since ${\bar H_X}$ has the deficiency indices $(N,N)$,  
${\bar H_X}$ has, in general, $N^2$-parameter family of 
self-adjoint extensions. 
All self-adjoint extensions $H_{U,X}$ of ${\bar H_X}$ are 
given by \cite{NE32,RS75}
\begin{eqnarray}
\label{eq2-11}
\lefteqn{D(H_{U,X})=\{
f+\sum\limits_{i=1}^N
{c_i (\psi_{i+}+\sum\limits_{j=1}^N {U_{ij}\psi_{j-}})} \vert 
f \in {\bar H_X}, c_i \in {\bf C} \},} \nonumber \\
& H_{U,X} \{ f+\sum\limits_{i=1}^N
{c_i (\psi_{i+}+\sum\limits_{j=1}^{N}{U_{ij}\psi_{j-}})} \} = 
{\bar H_X f} + i \Lambda \sum\limits_{i=1}^N
{c_i (\psi_{i+} - \sum\limits_{j=1}^{N}{U_{ij}\psi_{j-}})}, 
\end{eqnarray}
where $U_{ij}$ denotes a $N$-dimensional unitary matrix and 
\begin{eqnarray}
\label{eq2-12}
\psi_{j\pm}({\vec x})=\psi_j (\pm i\Lambda,{\vec x}), \ \ \Lambda > 0, 
\ \ \ j=1,\cdots,N, 
\end{eqnarray}
provide a basis for 
$Ker({\bar H_X}^* \mp i \Lambda)=
\overline{Ran(\bar{H}_X \pm i \Lambda)}^{\perp}$, 
respectively. 
Positive $\Lambda$ is regarded as a mass scale which can be 
arbitrarily fixed. 
The operator $H_{U,X}$ corresponds to the Hamiltonian for the system 
with $N$ point interactions. 
The special case $U=-1$ leads to the kinetic operator in $L^2 (S)$, 
$H_{-1,X}=H_0$, since $\psi_{j+}-\psi_{j-} \in D(H_0)$. 

By using Krein's formula,  
we can obtain the relation between two resolvents 
$(\omega - H_{0})^{-1}$ and $(\omega - H_{U,X})^{-1}$ for 
the unperturbed and perturbed systems respectively. 
Assume that $\bar{A}$ is a densely defined, closed symmetric operator 
in some Hilbert space with deficiency indices $(N,N)$. 
Let $B$ and $C$ be two self-adjoint extensions of $\bar{A}$ and denote 
by $\dot{A}$ the maximal common part of $B$ and $C$. 
Let $M$, $0<M \le N$, be the deficiency indices of $\dot{A}$ and 
let $\left\{ \phi_1(\omega),\cdots,\phi_M(\omega) \right\}$, 
which are linearly independent, span the corresponding deficiency 
subspace of $\dot{A}$; 
\begin{eqnarray}
\label{eq2-13}
\dot{A}^* \phi_i(\omega) = \omega \phi_i(\omega), 
\ \ \phi_i \in D(\dot{A}^* ), \ \ i=1,\cdots,M, 
\ \ Im \ \omega \neq 0.   
\end{eqnarray}
Then Krein's Formula reads \\[3mm]
{\bf Theorem } 
Let $B$, $C$, $\bar{A}$ and $\dot{A}$ be as above. 
Then 
\begin{eqnarray}
\label{eq2-14}
(\omega - B)^{-1} - (\omega - C)^{-1} =  
-\sum\limits_{i,j=1}^M {
\lambda_{ij}(\omega) \langle \phi_j (\bar{\omega}),\cdot 
\rangle \phi_i (\omega) }, 
\ \ \ \omega \in \rho(B) \cap \rho(C),  
\end{eqnarray}
where the matrix $\lambda(\omega)$ is nonsingular for 
$\omega \in \rho(B) \cap \rho(C)$ and 
$\lambda_{ij}(\omega)$ and $\phi_j(\omega)$, $i,j=1,\cdots,M$, 
may be chosen to be analytic in 
$\omega \in \rho(B) \cap \rho(C)$. 
Indeed, 
$\phi_i(\omega)$ can be defined as 
\begin{eqnarray}
\label{eq2-15}
\phi_i(\omega) = \phi_i(\omega_0) - (\omega-\omega_0)
(\omega-C)^{-1}\phi_i(\omega_0), \ \ \ i=1,\cdots,M, \ \ \ 
\omega \in \rho(C),  
\end{eqnarray}
where $\phi_i(\omega_0)$, $i=1,\cdots,M$, $Im \ \omega_0 \neq 0$, 
are linearly independent solutions of Eq.(\ref{eq2-13}) for 
$\omega=\omega_0$ and the inverse matrix of $\lambda(\omega)$ satisfies 
\begin{eqnarray}
\label{eq2-16}
\lambda_{ij}^{-1}(\omega)=\lambda_{ij}^{-1}(\omega')
-(\omega-\omega')
\langle \phi_j(\bar{\omega}), \phi_i(\omega') \rangle, 
\ \ \ i,j=1,\cdots,M, \ \ \omega,\omega' \in \rho(B) \cap \rho(C), 
\end{eqnarray}
if the $\phi_i(\omega)$, $i=1,\cdots,M$, are defined 
according to Eq.(\ref{eq2-15}). 
\\[3mm]
Applying Krein's formula ($\bar{A}=\bar{H}_X$,  
$B=H_{U,X}$, $C=H_0$ and hence $\dot{A}=\bar{H}_X$, $M=N$), 
we obtain the relation for $U \neq -1$, 
\begin{eqnarray}
\label{eq2-17}
(\omega - H_{U,X})^{-1} = (\omega - H_{0})^{-1} -  
\sum\limits_{i,j=1}^N {
\lambda_{ij}(\omega) \langle \psi_j (\bar{\omega}),\cdot 
\rangle \psi_i (\omega),   
}
\end{eqnarray}
where $\lambda^{-1}(\omega)$ satisfies  
\begin{eqnarray}
\label{eq2-18}
\lambda^{-1}_{ij}(\omega) - \lambda^{-1}_{ij}(\omega ') = 
-(\omega-\omega')  
\langle \psi_j(\bar{\omega}), \psi_i(\omega') \rangle 
= G_{ij}^{(0)}(\omega)-G_{ij}^{(0)}(\omega '), \ \ 
i,j=1,\cdots,N.   
\end{eqnarray}
Here we set 
\begin{eqnarray}
\label{eq2-19}
G_{ij}^{(0)}(\omega )\equiv G^{(0)}({\vec x_i},{\vec x_j};\omega )
= \sum\limits_{n=1}^\infty  
{{{\varphi _n({\vec x_i}) \varphi _n({\vec x_j})} 
\over {\omega -\varepsilon _n}}}.
\end{eqnarray}
The second equality in Eq.(\ref{eq2-18}) follows from 
the resolvent formula 
\begin{eqnarray}
\label{eq2-20}
(\omega - \omega')(\omega-H_0)^{-1}(\omega'-H_0)^{-1}=
(\omega'-H_0)^{-1}-(\omega-H_0)^{-1}.  
\end{eqnarray}
Each term on the right hand side in Eq.(\ref{eq2-18}) 
is a divergent series for $i=j$. However, the divergence cancels 
each other. 
The most important indication of Krein's formula is that  
it is sufficient to define $\lambda(\omega)$ 
for some fixed $\omega$ since then $\lambda(\omega)$ for any 
$\omega$ follows from Eq.(\ref{eq2-18}). 

The Green's function of $H_{U,X}$ is given by 
the coordinate representation of the resolvent Eq.(\ref{eq2-17});  
\begin{eqnarray}
\label{eq2-21}
G_{U,X}({\vec x},{\vec x'};\omega ) \equiv 
(\omega - H_{U,X})^{-1}({\vec x},{\vec x'}) 
= G^{(0)}({\vec x},{\vec x'};\omega ) -
\sum\limits_{i,j=1}^N {
G^{(0)}({\vec x},{\vec x_i};\omega )
\lambda_{ij}(\omega) G^{(0)}({\vec x_j},{\vec x'};\omega ). 
}
\end{eqnarray}
From Eq.(\ref{eq2-21}), we recognize that 
$- \lambda(\omega)$ corresponds to the transition matrix 
($T$-matrix) in the presence of $N$ point interactions. 
It follows from Eq.(\ref{eq2-11}) that 
\begin{eqnarray}
\label{eq2-22}
(\omega - H_{U,X}) (\psi_{i+}+\sum\limits_{j=1}^N {U_{ij}\psi_{j-}})=
(\omega - i \Lambda)\psi_{i+} + 
(\omega + i \Lambda) \sum\limits_{j=1}^N {U_{ij}\psi_{j-}}. 
\end{eqnarray}
Setting $\omega = -i \Lambda$ in Eq.(\ref{eq2-22}), we get 
\begin{eqnarray}
\label{eq2-23}
(-i \Lambda - H_{U,X})^{-1} \psi_{i+} =
-\frac{1}{2i \Lambda}(\psi_{i+}+\sum\limits_{j=1}^N {U_{ij}\psi_{j-}}). 
\end{eqnarray}
On the other hand, Eq.(\ref{eq2-17}) shows that the left hand side 
in Eq.(\ref{eq2-23}) is written as 
\begin{eqnarray}
\label{eq2-24}
(-i \Lambda - H_{U,X})^{-1} \psi_{i+} = 
(-i \Lambda - H_0)^{-1} \psi_{i+} - 
\sum\limits_{j,k=1}^N { \lambda_{jk} (-i \Lambda) 
\langle \psi_{k+}, \psi_{i+} \rangle \psi_{j-} }. 
\end{eqnarray}
The resolvent formula (\ref{eq2-20}) indicates 
\begin{eqnarray}
\label{eq2-25}
(-i \Lambda - H_0)^{-1} \psi_{i+} = 
-\frac{1}{2i \Lambda}(\psi_{i+} -\psi_{i-}), 
\end{eqnarray}
\begin{eqnarray}
\label{eq2-26}
\langle \psi_{k+},\psi_{i+} \rangle = 
\frac{1}{2i\Lambda } (
G_{ki}^{(0)}(-i\Lambda)-G_{ki}^{(0)}(i\Lambda)) 
= \frac{1}{2i\Lambda } (
\lambda^{-1}_{ki}(-i\Lambda) - 
\lambda^{-1}_{ki}(i\Lambda)),  
\end{eqnarray}
where the second equality in Eq.(\ref{eq2-26}) follows from Eq.(\ref{eq2-18}). 
Substituting Eqs.(\ref{eq2-25}) and (\ref{eq2-26}) into 
Eq.(\ref{eq2-24}) and compared with Eq.(\ref{eq2-23}), 
we obtain the relation between the unitary matrix 
and the $T$-matrix, 
\begin{eqnarray}
\label{eq2-27}
U_{ij} =  -\delta_{ij} +  
\sum\limits_{k=1}^N{ \lambda_{jk}(-i\Lambda)
(\lambda^{-1}_{ki}(-i\Lambda) -\lambda^{-1}_{ki}(i\Lambda)) } 
=  - \sum\limits_{k=1}^N{ 
\lambda_{jk}(-i\Lambda) \lambda^{-1}_{ki}(i\Lambda)}, 
\end{eqnarray}
or equivalently 
\begin{eqnarray}
\label{eq2-28}
U = -^{t}[\lambda(-i\Lambda) \lambda^{-1}(i\Lambda)]
\end{eqnarray}
in a simple matrix form.  
Noticing that Eq.(\ref{eq2-17}) implies 
\begin{eqnarray}
\label{eq2-29}
\lambda(\omega)^{\dagger} = \lambda(\bar{\omega}), 
\end{eqnarray}
we realize that the unitarity of $U$ is equivalent to the fact that 
$\lambda(i\Lambda)$ (resp. $\lambda(-i\Lambda)$) is a normal matrix. 
\section{The Case of Single Interaction}
In case of a single point interaction, 
a general form of $\lambda(\omega) \in {\bf C}$ 
which satisfies Eqs.(\ref{eq2-18}) and (\ref{eq2-29}) is given by  
\begin{eqnarray}
\label{eq3-1}
\lambda^{-1}(\omega) = {\bar G}(\omega)-{\bar v}^{-1}, 
\end{eqnarray}
where 
\begin{eqnarray}
\label{eq3-2}
{\bar G}(\omega) = 
\sum\limits_{n=1}^{\infty}{\varphi_n ({\vec x_1})^{2} (
\frac{1}{\omega-\varepsilon_n}+\frac{\varepsilon_n}{\varepsilon_n^2
+\Lambda^2})},  
\end{eqnarray}
\begin{eqnarray}
\label{eq3-3}
{\bar v}^{-1} = \Lambda \cot \frac{\theta}{2} 
\sum\limits_{n=1}^{\infty}{ \frac{\varphi_n ({\vec x_1})^{2}}
{\varepsilon_n^2+\Lambda^2}}, \ \ \ 0 \leq \theta < 2\pi.  
\end{eqnarray}
(Actually it would have been sufficient to define $\lambda(i\Lambda)$ 
since $\lambda(\omega)$ is calculated from Eq.(\ref{eq2-18}) with $N=1$, 
as mentioned in the previous section.) 
At first sight, the second term in Eq.(\ref{eq3-2}) seems to be 
redundant since it is energy-independent. 
However, it plays an essential role in making the problem well-defined; 
The infinite series in Eq.(\ref{eq3-2}) does not converge without 
the second term.
For spatial dimension $d\geq 4$, in general, 
the summation in Eq.(\ref{eq3-2}) diverges. 
This reflects the fact that 
the billiard problem with point interactions  
is not well-defined for $d\geq 4$ in quantum mechanics.
The energy independence of the counterterm also ensures that 
$\lambda(\omega)$ in Eq.(\ref{eq3-1}) satisfies Eq.(\ref{eq2-18}). 
(Note that, apart from the counterterm, ${\bar G}(\omega)$ 
has the same form as $G^{(0)}_{11}(\omega)$.)   
As one varies $0 \leq \theta < 2\pi$ for some fixed $\Lambda$, 
the value of $\bar{v}$, which is also energy-independent, 
ranges over the whole real number. 
We can formally consider $\bar{v}$ as a coupling strength 
of the point interaction. 
Its relation to physical strength of 
the interaction has been discussed elsewhere \cite{S94}. 
Here we just mention that 
the system approaches the empty billiard  
in the limit of $\bar{v} \rightarrow 0$. 
Inserting $\lambda(\omega)$ in Eq.(\ref{eq3-1}) 
together with Eqs.(\ref{eq3-2}) and (\ref{eq3-3}) into Eq.(\ref{eq2-28}), 
we get a simple expression of $U$ which manifests the unitarity; 
\begin{eqnarray}
\label{eq3-4}
U=- e^{i \theta}, 
\end{eqnarray}
irrespective to the value of $\Lambda$. 

The eigenvalues for the system perturbed by a single point interaction 
with coupling strength $\bar{v}$ located at $\vec{x}_1$ 
are obtained as the solutions of the equation 
\begin{eqnarray}
\label{eq3-5}
\lambda^{-1}(\omega) = 0, 
\end{eqnarray}
namely 
\begin{eqnarray}
\label{eq3-6}
{\bar G}(\omega)={\bar v}^{-1}.  
\end{eqnarray}
Eq.(\ref{eq3-2}) shows that 
within any interval between two neighboring unperturbed eigenvalues,
$\bar{G}(\omega)$ is a monotonically decreasing function 
that ranges over the whole real number.  This means that 
Eq.(\ref{eq3-6}) has a single solution on each interval; 
\begin{eqnarray}
\label{eq3-7}
\begin{array}{ll}
\displaystyle 
\varepsilon_1 < \omega_1 < 
\varepsilon_2 < \omega_2 < 
\varepsilon_3 < \omega_3 < \cdots\cdots, \ \ \ 
& 
\displaystyle 
{\rm for} \ \ \bar{v}>0, \\
\displaystyle 
\omega_1 < 
\varepsilon_1 < \omega_2 < 
\varepsilon_2 < \omega_3 < 
\varepsilon_3 < \cdots\cdots, \ \ \ 
& 
\displaystyle 
{\rm for} \ \ \bar{v}<0.  
\end{array}
\end{eqnarray}
The eigenfunction corresponding to an eigenvalue $\omega_{n}$ is written 
in terms of Green's function of the empty billiard as
\begin{eqnarray} 
\label{eq3-8} 
\psi_{n}(\vec{x}) =  
N_n G^{(0)}(\vec{x},\vec{x}_{1};\omega_{n}) 
=  
N_n \sum_{k=1}^{\infty} 
\frac{\varphi_{k}(\vec{x_{1}})}
     {\omega_{n}-\varepsilon_{k}}\varphi_{k}(\vec{x}), \ \ \ 
n=1,\cdots  
\end{eqnarray}
with a normalization constant 
\begin{eqnarray} 
\label{eq3-9} 
N_n = 1 \left/ \sqrt{\sum_{k=1}^{\infty} 
\left( \frac
{\varphi_k(\vec{x}_1)}{\omega_n-\varepsilon_k}\right)^2} 
\right., 
\ \ \ n=1,\cdots. 
\end{eqnarray}

In order to obtain each solution of Eq.(\ref{eq3-6}), 
a numerical task is needed in general. 
Our main purpose is, however, to examine under what conditions 
a point interaction brings about a significant effect 
in the statistical properties of spectrum. 
To achieve our aim without losing the essence, while still 
keeping loss of generality minimal, 
we introduce some approximations as follows. 
The first (and main) one is that the value of (the square of) 
each wave function at $\vec{x}_1$ is replaced by 
its average among many; 
\begin{eqnarray}
\label{eq3-10}
\varphi_n (\vec{x}_1)^2 \simeq 
\langle \varphi_n (\vec{x}_1)^2 \rangle
=1/S. 
\end{eqnarray}
Since it is often the case that the statistics 
are taken within a large number of, sometimes thousands of   
eigenstates (with $\vec{x}_1$ and $\bar{v}$ fixed), 
Eq.(\ref{eq3-10}) is quite satisfactory. 
Note that the second equality in Eq.(\ref{eq3-10}) is valid 
irrespective to the exact position of the interaction $\vec{x}_1$ 
as well as the energy region where the average is taken.  
Keeping Eq.(\ref{eq3-10}) in mind, 
we recognize from Eq.(\ref{eq3-8}) that 
if a perturbed eigenvalue $\omega_n$ is close to 
an unperturbed one $\varepsilon_n$ (or $\varepsilon_{n+1}$), 
then the corresponding eigenfunction $\psi_n$ 
is not substantially different from $\varphi_n$ (or $\varphi_{n+1}$).
Thus, the disturbance by a point interaction is mainly 
restricted to eigenstates with an eigenvalue around which 
$\bar{G}(\omega)$ has an inflection point.  
This is because each inflection point of $\bar{G}(\omega)$ is expected 
to appear, on average, around the midpoint on the interval between 
two neighboring unperturbed eigenvalues. 
Let $(\tilde{\omega},\bar{G}(\tilde{\omega}) )$ be one of 
such inflection points of $\bar{G}(\omega)$; 
$\ \ \tilde{\omega} \simeq (\varepsilon_{m}+\varepsilon_{m+1})/2 \ $ 
for some $m$.  
In this case, 
the contributions on $\bar{G}(\tilde{\omega})$ from 
the terms with $n \simeq m$ cancel each other, and 
we can approximate the summation in Eq.(\ref{eq3-2}) by a principal 
integral with a high degree of accuracy;   
\begin{eqnarray}
\label{eq3-11}
\bar{G}(\tilde{\omega}) & \simeq & \bar{g}(\tilde{\omega}), \\
\label{eq3-12}
\bar{g}(\omega) & = & 
\langle \varphi_{n}(\vec{x}_{1})^{2} \rangle \rho_{av} 
P  \int_{0}^{\infty} 
\left( \frac{1}{\omega-E}+\frac{E}{E^{2}+\Lambda^2} \right) dE, 
\end{eqnarray} 
where we have defined a continuous function $\bar{g}(\omega)$ 
which behaves like an interpolation of the inflection 
points of $\bar{G}(\omega)$. 
Clearly, $\bar{G}(\omega) \simeq \bar{g}(\omega)$ is valid only around 
the inflection points of $\bar{G}(\omega)$. 
Using an elementary indefinite integral 
\begin{eqnarray}
\label{eq3-13}
\int \left( \frac{1}{\omega-E}+\frac{E}{E^{2}+\Lambda^2} \right) dE  = 
-\ln \frac{ | \omega-E | }
{\sqrt{ E^2 + \Lambda^2}}, 
\end{eqnarray}
we obtain 
\begin{eqnarray}
\label{eq3-14}
\bar{G}(\tilde{\omega}) \simeq 
\langle \varphi_{n}(\vec{x}_{1})^{2} \rangle \rho_{av} 
\ln \frac{\tilde{\omega}}{\Lambda} 
\simeq \frac{M}{2\pi} 
\ln \frac{\tilde{\omega}}{\Lambda}, 
\end{eqnarray} 
where the second equality follows from 
Eqs.(\ref{eq2-3}) and (\ref{eq3-10}). 
Eq.(\ref{eq3-14}) indicates that the maximal coupling of 
a point interaction is attained with the coupling strength 
${\bar v}$ which satisfies 
\begin{eqnarray}
\label{eq3-15}
\bar{v}^{-1} \simeq \frac{M}{2\pi} \ln \frac{\omega}{\Lambda}. 
\end{eqnarray} 

The ``width'' of the strong coupling region 
(allowable error of $\bar{v}^{-1}$ in Eq.(\ref{eq3-15}))
can be estimated by considering a linearized eigenvalue equation. 
Expanding $\bar{G}(\omega)$ at $\omega=\tilde{\omega}$, 
we can rewrite Eq.(\ref{eq3-6}) as 
\begin{eqnarray}
\label{eq3-16}
\bar{G}(\tilde{\omega})+
\bar{G}'(\tilde{\omega})(\omega-\tilde{\omega}) \simeq \bar{v}^{-1}  
\end{eqnarray} 
or
\begin{eqnarray}
\label{eq3-17}
\bar{G}'(\tilde{\omega})(\omega-\tilde{\omega}) \simeq \bar{v}^{-1}  -
\frac{M}{2\pi} \ln \frac{\tilde{\omega}}{\Lambda}. 
\end{eqnarray} 
In order to ensure that the perturbed eigenvalue is 
close to $\tilde{\omega}$, the range of the right hand side 
in Eq.(\ref{eq3-17}) has to be restricted to 
\begin{eqnarray}
\label{eq3-18}
\left| \bar{v}^{-1}  - \frac{M}{2\pi} \ln \frac{\tilde{\omega}}{\Lambda}  
\right| \alt \frac{\delta_{\bar{v}^{-1}}}{2}
\end{eqnarray} 
where the width $\delta_{\bar{v}^{-1}}$ is defined by    
\begin{eqnarray}
\label{eq3-19}
\delta_{\bar{v}^{-1}} \equiv 
\left| \bar{G}'(\tilde{\omega}) \right| \rho_{av}^{-1}. 
\end{eqnarray} 
This is nothing but the average variance of the linearized $\bar{G}$ 
on the interval between the two 
unperturbed eigenvalues just below and above $\tilde{\omega}$.  
Using the approximation in Eq.(\ref{eq3-10}), 
the value of 
$\left| \bar{G}'(\tilde{\omega}) \right|$ 
can be estimated as follows;  
\begin{eqnarray}
\label{eq3-20}
\left| \bar{G}'(\tilde{\omega}) \right| 
= \sum_{n=1}^{\infty} \left(
\frac{\varphi_{n}(\vec{x}_{1})}
{\tilde{\omega}-\varepsilon_{n}} \right)^2
\simeq \langle \varphi_{n}(\vec{x}_{1})^{2} \rangle 
\sum_{n=1}^{\infty} 
\frac{2}
{\{(n-\frac{1}{2})\rho_{av}^{-1}\}^2} 
=  
8 \langle \varphi_{n}(\vec{x}_{1})^{2} \rangle 
\rho_{av}^2 
\sum_{n=1}^{\infty} \frac{1}{(2n-1)^2} 
= \pi^{2} \langle \varphi_{n}(\vec{x}_{1})^{2} \rangle \rho_{av}^2. 
\end{eqnarray}
The second equality follows from the approximation that 
the unperturbed eigenvalues are distributed with a mean 
interval $\rho_{av}^{-1}$ in the whole energy region. 
This assumption is quite satisfactory, 
since the denominator of $\bar{G}'(\omega)$ is of the order of 
$(\omega-\varepsilon_n)^2$, indicating that the summation in 
Eq.(\ref{eq3-20}) converges rapidly. 
From Eq.(\ref{eq3-20}), we obtain 
\begin{eqnarray}
\label{eq3-21}
\delta_{\bar{v}^{-1}} \simeq 
\pi^{2} \langle \varphi_{n}(\vec{x}_{1})^{2} \rangle \rho_{av} 
\simeq \frac{\pi M}{2}. 
\end{eqnarray} 
From Eqs.(\ref{eq3-18}) and (\ref{eq3-21}), we recognize that 
the effect of a point interaction of coupling strength $\bar{v}$ 
on statistical properties of spectrum is 
substantial only in the eigenstates with eigenvalue $\omega$ 
such that 
\begin{eqnarray}
\label{eq3-22}
\left| \bar{v}^{-1} - \frac{M}{2\pi} \ln \frac{\omega}{\Lambda}  \right| 
& \alt & \frac{\delta_{\bar{v}^{-1}}}{2}  \simeq  
\frac{\pi M}{4} \simeq M. 
\end{eqnarray}
 
We do not go into details of numerical experiments in this paper, 
but just mention that the validity of the conjecture (\ref{eq3-22}) 
has been confirmed by examining the spectrum of rectangular billiards 
with a single point interaction inside \cite{S94}. 
Concerning the nearest-neighbor level spacing distribution $P(S)$, 
a level repulsion seen in Eq.(\ref{eq1-2}) is observed under the 
condition (\ref{eq3-22}), while $P(S)$ is not substantially different 
from the Poisson prediction, Eq.(\ref{eq1-1}), otherwise. 

Before closing this section, we give a short comment  
on the shape of the billiard. 
Our implicit assumption for the shape is that 
the average level density of the empty billiard  
is dominated by the area term, which does not 
depend on energy. 
The assumption is justified for a generic 
two-dimensional billiard which has 
the same order of length scale in each direction,      
irrespective to a full detail of the shape of the billiard. 

\section{The Case of Multiple Interactions}
We intend to generalize Eq.(\ref{eq3-1}) to finitely many point 
interactions. This is achieved by introducing the $N$-parameter 
family of self-adjoint extensions of ${\bar H_X}$ defined 
by the unitary matrix $U$, Eq.(\ref{eq2-28}), together with  
\begin{eqnarray}
\label{eq4-1}
\lambda^{-1}_{ij}(\omega) = 
\left\{ 
\begin{array}{lll}
 {\bar G}_{i}(\omega) - {\bar v}^{-1}_{i}, & 
{\rm for} \ \  i=j, & \ \  {\bar v}_{i} \in {\bf R}, \\
G^{(0)}_{ij}(\omega), & {\rm for} \ \  i \neq j, & 
\end{array}\right.
\ \ \ i,j = 1,\cdots,N, 
\end{eqnarray}
where $G^{(0)}_{ij}(\omega)$ is defined in Eq.(\ref{eq2-19}) and 
\begin{eqnarray}
\label{eq4-2}
{\bar G}_{i}(\omega) = 
\sum\limits_{n=1}^{\infty}{\varphi_n ({\vec x_i})^{2} (
\frac{1}{\omega-\varepsilon_n}+\frac{\varepsilon_n}{\varepsilon_n^2
+\Lambda^2})}, \ \ \ i=1,\cdots,N. 
\end{eqnarray}
Eq.(\ref{eq4-1}) with opposite sign 
describes the $T$-matrix of the system with 
$N$ point interactions which satisfy separated boundary 
conditions at each point $\vec{x}_{i}$.  
The coupling strength of the $i$-th interaction is 
assigned by a real number ${\bar v}_i$, $i=1,\cdots,N$. 
The eigenvalues of this system are the poles of the $T$-matrix, 
determined by 
\begin{eqnarray}
\label{eq4-3}
\det \ \lambda^{-1}(\omega)=0. 
\end{eqnarray}
Qualitative behavior 
of the eigenvalues of $\lambda^{-1}(\omega)$ as a function of $\omega$ 
can be examined as follows. 
Let us cut the infinite series of each matrix element of 
$\lambda^{-1}(\omega)$ at $n_{max}$ which satisfies 
$\omega \ll \epsilon_{n_{max}}$.
We then have 
\begin{eqnarray}
\label{eq4-4}
\lambda^{-1}(\omega) = 
\lim_{n_{max}\rightarrow\infty} \Sigma^{(0)}(\omega)
\end{eqnarray} 
where $\Sigma^{(0)}(\omega)$ is a sum of an ``unperturbed part'' $T^{(0)}$ 
and an ``interaction part'' $V^{(0)}(\omega)$;  
\begin{eqnarray}
\label{eq4-5}
\Sigma^{(0)}(\omega) & = & T^{(0)} + V^{(0)}(\omega)
\end{eqnarray} 
with 
\begin{eqnarray}
\label{eq4-6}
T^{(0)} = \left ( 
\begin{array}{cccc}
d_1^{(0)} & & & \bigzerou \\ 
& d_2^{(0)} & & \\ 
& & \ddots & \\
\bigzerol & & & d_N^{(0)} 
\end{array}
\right ), \ \ \ \ \ 
V^{(0)}(\omega) = 
\sum_{n=1}^{n_{max}} \frac{1}{\omega -\varepsilon_n} 
\left ( 
\begin{array}{c}
\varphi_n^{(0)}(\vec{x}_1) \\
\vdots \\
\vdots \\
\varphi_n^{(0)}(\vec{x}_N) \\
\end{array}
\right ) 
\left ( 
\begin{array}{cccc}
\varphi_n^{(0)}(\vec{x}_1) & \cdots &
\varphi_n^{(0)}(\vec{x}_N) 
\end{array}
\right ). 
\end{eqnarray} 
Here we set 
\begin{eqnarray}
\label{eq4-8}
d_i^{(0)} = \sum_{n=1}^{n_{max}} \varphi_n (\vec{x}_i)^2 
\frac{\varepsilon_n}{\varepsilon_n^2 + \Lambda^2} - \bar{v}_i^{-1}, 
\ \ \ \ 
\varphi_n^{(0)}(\vec{x}_i) & = & \varphi_n(\vec{x}_i), 
\ \ \ \ i=1,\cdots,N. 
\end{eqnarray}
The $N$-dimensional matrix $\Sigma^{(0)}(\omega)$ formally has a form 
similar to the Hamiltonian with $n_{max}$ rank-one (separable) 
interactions. We realize that 
$d_i^{(0)}$ takes a role of the ``unperturbed energy'',  
while the inverse of $\omega-\varepsilon_n$ corresponds to 
the ``strength'' of the $n$-th interaction. 
In the following, we assume that 
$d_1^{(0)} < d_2^{(0)} < \cdots < d_N^{(0)}$ without a loss of generality. 

The eigenvalues of $\Sigma^{(0)}(\omega)$ can be obtained in a perturbative 
manner as follows. 
We start by setting $k=1$ and define 
\begin{eqnarray}
\label{eq4-10}
\tilde{T}^{(k)}(\omega) = T^{(k-1)}(\omega) + 
\frac{1}{\omega -\varepsilon_k} 
\left ( 
\begin{array}{c}
\varphi_k^{(k-1)}(\vec{x}_1) \\
\vdots \\
\vdots \\
\varphi_k^{(k-1)}(\vec{x}_N) \\
\end{array}
\right ) 
\left ( 
\begin{array}{cccc}
\varphi_k^{(k-1)}(\vec{x}_1) & \cdots &
\varphi_k^{(k-1)}(\vec{x}_N) 
\end{array}
\right ).   
\end{eqnarray}
Namely, only the first term in the summation of $V^{(k-1)}(\omega)$ is 
taken into account in the matrix $\tilde{T}^{(k)}(\omega)$. 
Since $\tilde{T}^{(k)}(\omega)$ has a single rank-one interaction, 
we obtain the eigenvalues of $\tilde{T}^{(k)}(\omega)$ 
as the solutions $d_i^{(k)}$, $i=1,\cdots,N$, of the equation 
\begin{eqnarray}
\label{eq4-11}
\sum_{i=1}^{N} \frac{\varphi_k^{(k-1)}(\vec{x}_i)^2}{d^{(k)}-d_i^{(k-1)}}
= \omega-\varepsilon_k. 
\end{eqnarray} 
It is worthy to note that 
\begin{eqnarray}
\label{eq4-11b}
d_{i-1}^{(k-1)}<d_i^{(k)}<d_{i}^{(k-1)}  \ \ \ 
{\rm for} \ \ \omega < \varepsilon_k, \ \ \ \ \  
d_{i}^{(k-1)}<d_i^{(k)}<d_{i+1}^{(k-1)} \ \ \ 
{\rm for} \ \  \omega > \varepsilon_k.    
\end{eqnarray}
We define an orthogonal transformation $\Omega^{(k)}(\omega)$ such that 
the matrix $\tilde{T}^{(k)}(\omega)$ is diagonal in the new basis,  
\begin{eqnarray}
\label{eq4-12}
T^{(k)}(\omega) \equiv  ^{t} 
\! \! \Omega^{(k)}(\omega)\tilde{T}^{(k)}(\omega)\Omega^{(k)}(\omega) 
= \left ( 
\begin{array}{cccc}
d_1^{(k)} & & & \bigzerou \\ 
& d_2^{(k)} & & \\ 
& & \ddots & \\
\bigzerol & & & d_N^{(k)} 
\end{array}
\right ). 
\end{eqnarray}
Clearly, the $i$-th column of $\Omega^{(k)}(\omega)$  
is given by the normalized 
eigenfunction corresponding to the $i$-th eigenvalue $d_i^{(k)}$, 
the exact form of which is unnecessary in the present argument. 
With the new basis, the new Hamiltonian matrix has a form  
\begin{eqnarray}
\label{eq4-13}
\Sigma^{(k)}(\omega)
 \equiv  
^{t} \! \! \Omega^{(k)}(\omega)\Sigma^{(k-1)}(\omega)\Omega^{(k)}(\omega) 
=T^{(k)}(\omega)+V^{(k)}(\omega), 
\end{eqnarray}
where 
\begin{eqnarray}
\label{eq4-14}
V^{(k)}(\omega) =
\sum_{n=k+1}^{n_{max}} \frac{1}{\omega -\varepsilon_n} 
\left ( 
\begin{array}{c}
\varphi_n^{(k)}(\vec{x}_1) \\
\vdots \\
\vdots \\
\varphi_n^{(k)}(\vec{x}_N) \\
\end{array}
\right ) 
\left ( 
\begin{array}{cccc}
\varphi_n^{(k)}(\vec{x}_1) & \cdots &
\varphi_n^{(k)}(\vec{x}_N) 
\end{array}
\right ) 
\end{eqnarray} 
together with 
\begin{eqnarray}
\label{eq4-15}
\left ( 
\begin{array}{cccc}
\varphi_n^{(k)}(\vec{x}_1) & \cdots &
\varphi_n^{(k)}(\vec{x}_N) 
\end{array}
\right ) =
\left ( 
\begin{array}{cccc}
\varphi_n^{(k-1)}(\vec{x}_1) & \cdots &
\varphi_n^{(k-1)}(\vec{x}_N) 
\end{array} 
\right ) \Omega^{(k)}(\omega),   \ \ \ \ 
n=k+1,\cdots,n_{max}. 
\end{eqnarray} 
The new interaction matrix $V^{(k)}(\omega)$ keeps the separability of the 
interaction, while the number of the terms decreases by one, 
compared to $V^{(k-1)}(\omega)$. 
The procedure mentioned above can be repeated for 
$k=2,\cdots,n_{max}$, successively. 
After the repetition, 
we obtain $d_i^{(n_{max})}$, $i=1,\cdots,N$, which are exactly 
the eigenvalues of $\Sigma^{(0)}(\omega)$. 

The above algorithm serves to understand the qualitative behavior 
of the eigenvalues of $\lambda^{-1}(\omega)$. 
The eigenvalue equation (\ref{eq4-11}) indicates that  
if $\varepsilon_{k}$ is far from the energy 
$\omega$ under consideration, the ``strength'' of the rank-one 
interaction, $(\omega-\varepsilon_k)^{-1}$, is very weak and it has 
little effect except that each eigenvalue 
shifts by 
\begin{eqnarray}
\label{eq4-16}
d_i^{(k)} \simeq d_i^{(k-1)} + 
\frac{\varphi_k^{(k-1)}(\vec{x}_i)^2}{\omega-\varepsilon_k} 
\simeq d_i^{(k-1)} + 
\frac{1}{S(\omega-\varepsilon_k)}. 
\end{eqnarray}
This means that $\Omega^{(k)}(\omega)$ substantially differs from 
the unit matrix only in case of $\varepsilon_{k} \simeq \omega$.  
Thus, we can approximate $\Sigma^{(0)}(\omega)$ by 
\begin{eqnarray}
\label{eq4-17}
\Sigma^{(0)}(\omega) & \simeq & \bar{T}(\omega) + \bar{V}(\omega) 
\end{eqnarray}
with
\begin{eqnarray}
\label{eq4-18}
\bar{T}(\omega) = \left ( 
\begin{array}{cccc}
\bar{d}_1 & & & \bigzerou \\ 
& \bar{d}_2 & & \\ 
& & \ddots & \\
\bigzerol & & & \bar{d}_N 
\end{array}
\right ), \ \ \ \ \ 
\bar{V}(\omega) =  
\sum_{\epsilon_n \simeq \omega} \frac{1}{\omega -\varepsilon_n} 
\left ( 
\begin{array}{c}
\varphi_n (\vec{x}_1) \\
\vdots \\
\vdots \\
\varphi_n (\vec{x}_N) \\
\end{array}
\right ) 
\left ( 
\begin{array}{cccc}
\varphi_n (\vec{x}_1) & \cdots &
\varphi_n (\vec{x}_N) 
\end{array}
\right ). 
\end{eqnarray} 
Here we set 
\begin{eqnarray}
\label{eq4-20}
\bar{d}_i & = & \sum_{n=1}^{n_{max}} \varphi_n (\vec{x}_i)^2 
\frac{\varepsilon_n}{\varepsilon_n^2 + \Lambda^2} 
+ \sum_{n=1 \atop \varepsilon_n \simeq \! \! \! \! \setminus \omega}^{n_{max}} 
\frac{\varphi_n (\vec{x}_i)^2}{\omega-\epsilon_n} - \bar{v}_i^{-1}, \ \ \ \ 
i=1,\cdots,N. 
\end{eqnarray}
Note that $\bar{d}_i$ shows the tracks of the discard terms 
in $\bar{V}(\omega)$.  
In the limit of $n_{max}\rightarrow \infty$, 
we can estimate 
\begin{eqnarray}
\label{eq4-21}
\bar{d}_i \simeq \frac{M}{2\pi} \ln \frac{\omega}{\Lambda}-\bar{v}_i^{-1} 
\end{eqnarray}
as before. 
Since the shift of eigenvalues caused by rank-one interactions is limited 
[see Eq.(\ref{eq4-11b})], 
we expect from Eq.(\ref{eq4-18}) together with Eq.(\ref{eq4-21}) that 
the condition (\ref{eq3-22}) for a single point interaction 
can be generalized to the case of multiple interactions. 
Namely, the $i$-th interaction with strength $\bar{v}_i$ 
affects the spectral properties in the energy region which satisfies 
\begin{eqnarray}
\label{eq4-22}
\left| \bar{v}_{i}^{-1} - \frac{M}{2\pi} \ln \frac{\omega}{\Lambda}  \right| 
\alt \frac{\pi M}{4} \simeq M. 
\end{eqnarray}

\section{Conclusion}

We have discussed the spectral properties 
of two-dimensional Laplacian with a finite number of 
point interactions. 
In spite of the apparent simplicity, 
careful treatments are required for point interactions. 
Based on the self-adjoint extension theory in functional analysis, 
we have deduced the Green's function appropriate for the operator. 
The problem is based on obvious physical motivations. 
The operator exactly corresponds to the quantum-mechanical 
Hamiltonian for a point particle moving in 
a two-dimensional bounded region (billiard) with point impurities 
inside. 
Based on a general argument, 
we have clarified the condition under which 
the statistical properties of spectrum are substantially 
influenced by the point interactions. 
The findings are summarized as follows;\\[3mm]   
(1) \hspace{0.5ex} 
For a two-dimensional billiard, the effect of a point interaction 
with coupling strength $\bar{v}$ on statistical properties of spectrum 
is maximal under the condition 
\begin{eqnarray}
\label{eq5-1}
\bar{v}^{-1} \simeq \frac{M}{2\pi}\ln \frac{\omega}{\Lambda}, 
\end{eqnarray}
where $M$ is the mass of a particle moving in the billiard and 
$\Lambda$ is an arbitrary mass scale. This indicates that 
the maximal coupling region shifts with a logarithmic dependence 
of the energy $\omega$ in two dimension. \\[3mm]   
(2) \hspace{0.5ex} 
The width $\delta_{\bar{v}^{-1}}$ (or an allowable error in $\bar{v}^{-1}$ 
to look for the effect) is estimated as 
\begin{eqnarray}
\label{eq5-2}
\delta_{\bar{v}^{-1}} \simeq \frac{\pi M}{2}, \nonumber \\
\end{eqnarray}
which is energy-independent. If the value of $\bar{v}^{-1}$ differs 
from the right hand side in Eq.(\ref{eq5-1}) 
to the extent of Eq.(\ref{eq5-2}) at energy $\omega$, 
the effect of the point interaction on the statistics tends to disappear.  
\\[3mm]   
(3) \hspace{0.5ex} 
(1) and (2) are generalized to the case of multiple point 
interactions under separated boundary conditions. 
If we collect the interactions with the same order of magnitude of the 
coupling strength as a single group, we expect that the interactions 
belonging to one of such groups disturb the particle 
motion in a ``coherent'' manner in the energy region determined 
by Eq.(\ref{eq5-1}) \cite{CS96}, while their influence hardly appears 
in different energy regions where the difference between both sides 
in Eq.(\ref{eq5-1}) is larger than Eq.(\ref{eq5-2}).

\end{document}